\documentclass[12pt,a4paper]{article}
\usepackage{amsmath}
\usepackage{amssymb}
\usepackage{caption}
\usepackage{fancyhdr}
\usepackage{fancybox}
\usepackage{graphicx}
\usepackage{tikz}
\usepackage[final,color]{showkeys}
\usepackage[english]{babel}

\newcommand{\bbf}{\mathbf{f}}
\newcommand{\bu}{\mathbf{u}}

\renewcommand\baselinestretch{1.12}

\begin{document}

\begin{center}
\textbf{\Large Propagation of nonlinear waves \\[1mm] in a rarefied bubbly flow} \\[2mm]
\large 
Alexander A. Chesnokov$^{1,2}$, Maxim V. Pavlov$^{3,4}$ \\[2mm]
$^{1}$Lavrentyev Institute of Hydrodynamics SB RAS \\
15 Lavrentyev Ave., Novosibirsk, 630090, Russia \\ 
chesnokov@hydro.nsc.ru \\[2mm]

$^{2}$ Novosibirsk State University \\
2 Pirogova Str., Novosibirsk, 630090, Russia \\[2mm]

$^{3}$ Lebedev Physical Institute of Russian Academy of Sciences \\
Leninskij Prospekt 53, 119991 Moscow, Russia \\
m.v.pavlov@lboro.ac.uk \\[2mm]

$^4$ Laboratory of Geometrical Methods in Mathematical Physics, \\
Lomonosov Moscow State University, \\
Leninskie Gory 1, 119991 Moscow, Russia\\[2mm]

\end{center}

\normalsize

\tableofcontents

\newpage

\begin{abstract}
The one-dimension Russo--Smereka kinetic equation describing the propaga\-tion of nonlinear concentration waves in a rarefied bubbly fluid is considered. Reductions of the model to finite component systems are derived. Stability of the bubbly flow in terms of hyperbolicity of the kinetic equation is studied. Conservation form of the model is proposed and numerical solution of the Cauchy problem with discontinuous initial data is obtained.
\end{abstract}


\vspace{2mm}

\section{Introduction}

A kinetic theory based on the statistical description of the interac\-tion of a large number of bubbles has been developed for modelling of nonlinear waves in a rarefied bubbly flow \cite{RS96a, Herrero99, TeshGavr02}. The kinetic models taking into account the effect of collective interaction
between bubbles are derived using the system of Hamilton's ODEs describing the motion of individual bubbles. To obtain this system of equations, one needs to know the kinetic energy of the fluid \cite{MilneTh60}. Assuming that all bubbles are rigid massless spheres of the same radius, Russo and Smereka \cite{RS96a} approximately calculated the energy and Hamiltonian of a 
bubble motion and obtained the kinetic equation for the evolution of the one-particle distribution function. This model is analogous to the Vlasov equation for plasma flow and describes the collective behavior of a large number of bubbles subject to long-range interactions, modelled by
self-consistent field.

The characteristic properties of the Russo--Smereka kinetic equation for the case of one space variable are studied by Teshukov \cite{Tesh99} on the base of a generalized theory of characteristics and notion of hyperbolicity for integro-differential equations \cite{Tesh85, LT00}. In \cite{Tesh99}, hyperbolicity conditions of the model are formulated and Riemann invariants and infinite series of conservation laws are found. The exact solutions of
the Russo--Smereka kinetic equation in the classes of travelling and simple waves, as well as solutions with linearly dependent Riemann invariants, are obtained and studied in \cite{Chesn00, RTC05}.

In the paper, we propose a conservation form of the kinetic model, which allows one to consider discontinuous solutions. Differential conservation laws approximating the Russo--Smereka kinetic equation are derived. These laws are used to perform numerical calculations of wave propagation in a rarefied bubbly flow, which show the possibility of the kinetic roll-over (the formation of two peaks of the distribution function which originally had a single peak). Stability analysis of the flows in terms of hyperbolicity of the model shows that the effect of the kinetic roll-over of the distribution function leads to instability of the flow. Some reductions of the kinetic equations to finite component systems are also obtained and their properties are studied.

The structure of the paper is as follows. In Section 2 we present different formulations of the Russo--Smereka kinetic model. Each formulation more suitable for some specific application. We also recall hyperbolicity conditions for this integro-differential model and give an example of verification of the hyperbolicity conditions. Conservation form of the kinetic model for a bubbly flow is proposed in Sections 3, as well as its approximation on the base of a system of differential conservation laws with a large number of unknowns. In Section 4, we introduce the hydrodynamic chain associated with the Russo--Smereka kinetic equation. We show that this chain possesses infinitely many conservation laws. This means that this chain as well as the Russo--Smereka kinetic equation are integrable systems. Then we constructed finite-component reductions utilizing generalized functions. In Section 5 we proved that the hydrodynamic chain associated with the Russo--Smereka kinetic model is the modified Benney hydrodynamic chain derived by Kupershmidt. We utilized the method of hydrodynamic reductions, which allows to construct infinitely many particular solutions for the Russo--Smereka kinetic model. Numerical results we present in Section 6. Finally, we draw some conclusions.  

\section{Mathematical model}

In the one-dimensional case, the Russo--Smereka kinetic equation in dimensionless variables is written as follows \cite{RS96a, Tesh99} 
\begin{equation}  \label{eq:RS-1D}
  \frac{\partial f}{\partial t} + (p-j) \frac{\partial f}{\partial x} + p 
  \frac{\partial j}{\partial x} \frac{\partial f}{\partial p}=0, \quad
  j(t,x)=\int pf(t,x,p)\,dp.
\end{equation}
Here $f(t,x,p)\geq 0$ is the distribution function for bubbles in phase space, $x$ and $p$ are the position and momentum of the bubble, $t$ is time, and $j$ is the first moment of the distribution function. It is assumed that $f$ decreases rapidly at infinity or is finite over the variable $p$. The model is suitable for the description of a rarefied bubbly flow in the case
of small pressure variations. The condition for a bubbly flow to be rarefied is given by the inequality 
\begin{equation} \label{eq:n} 
  n(t,x)=\int f\, dp <1. 
\end{equation}

As was shown in \cite{Tesh99}, to study the properties of the kinetic equation \eqref{eq:RS-1D}, it is appropriate to transform to the Eulerian--Lagrangian coordinates $x$ and $\lambda$ by substitution the variable $p=p(t,x,\lambda)$, where the function $p(t,x,\lambda)$ is a
solution of the Cauchy problem 
\begin{equation}  \label{eq:semi-L}
  p_t+(p-j)p_x=pj_x, \quad p(0,x,\lambda)=p_0(x,\lambda), 
  \quad \lambda\in [0,1].
\end{equation}
As result, for the new desired functions $p(t,x,\lambda)$ and $f(t,x,\lambda) $, we obtain the integro-differential system of equations \cite{Tesh99, LT00}
\begin{equation}  \label{eq:RS-Larg}
 \begin{array}{l} \displaystyle 
   p_t+(p-j)p_x-pj_x=0, \quad f_t+(p-j)f_x=0, \quad 
   j=\int p p_\lambda f\,d\lambda.
 \end{array}
\end{equation}

Indeed, let us show that Eqs.\,\eqref{eq:RS-Larg} are derived as a consequence of the Russo--Smereka model \eqref{eq:RS-1D}. We use $\tilde{f}$ to denote the distribution function in semi-Lagrangian coordinates: 
\[ \tilde{f}(t,x,\lambda)=f(t,x,p(t,x,\lambda)). \]
Then, the derivatives are represented as 
\[ \frac{\partial \tilde{f}}{\partial t}= \frac{\partial f}{\partial t}+  
   \frac{\partial f}{\partial p}\frac{\partial p}{\partial t}, \quad  
   \frac{\partial \tilde{f}}{\partial x}= \frac{\partial f}{\partial x}+  
   \frac{\partial f}{\partial p}\frac{\partial p}{\partial x}\,. \]
These formulae and \eqref{eq:semi-L} imply the obvious equalities 
\[ \frac{\partial \tilde{f}}{\partial t}+(p-j)
   \frac{\partial \tilde{f}}{\partial x}=  \frac{\partial f}{\partial t}+
   (p-j)\frac{\partial f}{\partial x}+  p\frac{\partial j}{\partial x} 
   \frac{\partial f}{\partial p}=0 \]
which lead to the second equation of system \eqref{eq:RS-Larg}.

In what follows, we assume that the inequality $p_{\lambda }>0$ is satisfied, which provides invertibility of the change of variables.

\subsection{Hyperbolicity conditions for the kinetic equation}

Eqs.\,\eqref{eq:RS-Larg} belong to the class of systems with operator coefficients of the form 
\begin{equation}  \label{eq:Absr-Eq}
  \mathbf{U}_t+\mathbf{A}\langle\mathbf{U}_x\rangle =0,
\end{equation}
for which a generalization of the hyperbolicity notion was proposed in~\cite{Tesh85}. Here, $\mathbf{U}(t,x,\lambda)$ is unknown vector function and $\mathbf{A}$ is a non-local operator on a set of functions of $\lambda$. Characteristics of the system~\eqref{eq:Absr-Eq} are determined by equation $x'(t)=k(t,x)$, where $k$ is an eigenvalue of problem $(\mathbf{F}, (\mathbf{A}-kI)\langle\mbox{\boldmath$\varphi$}\rangle)=0$. The eigenfunctional $\mathbf{F}$ is defined on a set of functions of variable~$\lambda$, while the values of $t$ and $x$ are considered to be fixed, and is sought in a class of locally integrable or generalized functions ($\mbox{\boldmath$\varphi$}(\lambda)$ is a smooth test function). The system of eqs.\,\eqref{eq:Absr-Eq} is hyperbolic if all eigenvalues $k$ are real and the set of relations on the characteristics $(\mathbf{F},\mathbf{U}_t+k\mathbf{U}_x)=0$ is equivalent to \eqref{eq:Absr-Eq}.

As it follows from~\cite{Tesh99, LT00}, hyperbolicity conditions for the Russo--Smereka kinetic equation~\eqref{eq:RS-1D} on a solution $f(t,x,p)$ are formulated in terms of characteristic function 
\begin{equation}  \label{eq:chi-f}
  \chi(k+j)=1-n+(k+j)^2\int\frac{f\,dp}{(p-k-j)^2}\,,
\end{equation}
or, more precisely, in terms of its limit values on the real axis from upper and lower complex half-planes 
\begin{equation}  \label{eq:chi-pm}
  \chi^\pm(p)=1-n(t,x)+p^2\int\frac{\partial f(t,x,p')}{\partial p'}\frac{dp'}{p'-p} 
  \pm \pi i p^2 \frac{\partial f(t,x,p)}{\partial p} \,,
\end{equation}
which are obtained from \eqref{eq:chi-f} by integration by parts and application of Sokhotski--Plemelj formulae.

According to~\cite{Tesh99, LT00}, kinetic equation~\eqref{eq:RS-1D} is hyperbolic on the rapidly decreasing solution $f(t,x,p)$ if the following conditions hold 
\begin{equation}  \label{eq:hyp-cond}
  \Delta \mathrm{arg} \chi^\pm(p)=0.
\end{equation}
The argument increment is calculated when $p$ changes from $-\infty$ to $\infty$ at fixed values of variables $t$ and $x$. If $\mathrm{supp}\,f$ is bounded, then the conditions \eqref{eq:hyp-cond} take the form 
\begin{equation}  \label{eq:hyp-cond-1}
  \Delta \mathrm{arg} \big(\chi^+(p)/\chi^-(p)\big)=0,
\end{equation}
where the argument increment is calculated for $p \in \mathrm{supp}\,f$. The hyperbolicity conditions \eqref{eq:hyp-cond} (or \eqref{eq:hyp-cond-1}) guarantee that characteristic equation $\chi(k+j)=0$ has no complex roots, and these conditions are necessary for flow stability.

\subsection{Example of verification of the hyperbolicity conditions}

Conditions~\eqref{eq:hyp-cond} (or \eqref{eq:hyp-cond-1}) allow one to verify whether the Russo--Smereka kinetic equation~\eqref{eq:RS-1D} is hyperbolic for a given solution $f(t,x,p)$. Following \cite{LT00, KhCh11}, in the plane $(Z^1,Z^2)$ we construct a closed contour $C$ consisting of the contours $C^+$ and $C^-$. The contour $C^+$ is given parametrically by the
equations 
\[ Z^1=\mathrm{Re} \{\chi^+(p)\}, \quad Z^2=\mathrm{Im} \{\chi^+(p)\}. \]
where the complex functions $\chi^\pm(p)$ are defined by \eqref{eq:chi-pm}. A contour $C^-$, which is symmetric about the $Z^1$ axis to the contour $C^+$ is given by the same equation with the function $\chi^-(p)$. If the point of $Z^1=0$, $Z^2=0$ lies in the domain bounded by the contour $C$, then the characteristic equation $\chi(k)=0$ has complex roots (function $\chi(k)$ is
given by \eqref{eq:chi-f}). Otherwise, the kinetic equation for the corresponding solution is hyperbolic.

In the theory of plasma waves the following result is known \cite{Stix}: any solution of the one-dimensional linearized Vlasov equation is stable if it defined by a distribution function with a single maximum. We show that in the kinetic theory for a bubbly flow it is not true. As mentioned above, hyperbolicity conditions~\eqref{eq:hyp-cond} are violated if the point $Z^1=0$, $Z^2=0$ is in the domain bounded by the contour $C$. In view of the inequality $Z^1 \to 1-n>0$ for $p\to \pm \infty$ (here $n$ is given by \eqref{eq:n}), this is possible only if the conditions $Z^1(p_*)<0$ and $Z^2(p_*)=0$ are satisfied at some point $p_* \in (-\infty,\infty)$. The equality $Z^2(p_*)=0$ is satisfied at
the single interior point of $p=p_c$ at which the distribution function $f(p) $ reaches a local maximum. Since the distributions $f=f(p)$ with one maximum obey the inequality $(p_c-p)f'(p)\geq 0$, one can conclude that the sign of the quantity 
\[ Z^1(p_c)=1-n+p_c^2\int\frac{f^{\prime }(p)dp}{p-p_c} \]
depends on the point $p_c$. Bubbly flow is stable if the extreme point $p_c$ is sufficiently close to zero (this guarantee that the inequality $Z^1(p_c)>0$ is fulfilled). 

\begin{figure}[htb]
\begin{center}
\resizebox{1\textwidth}{!}{\includegraphics{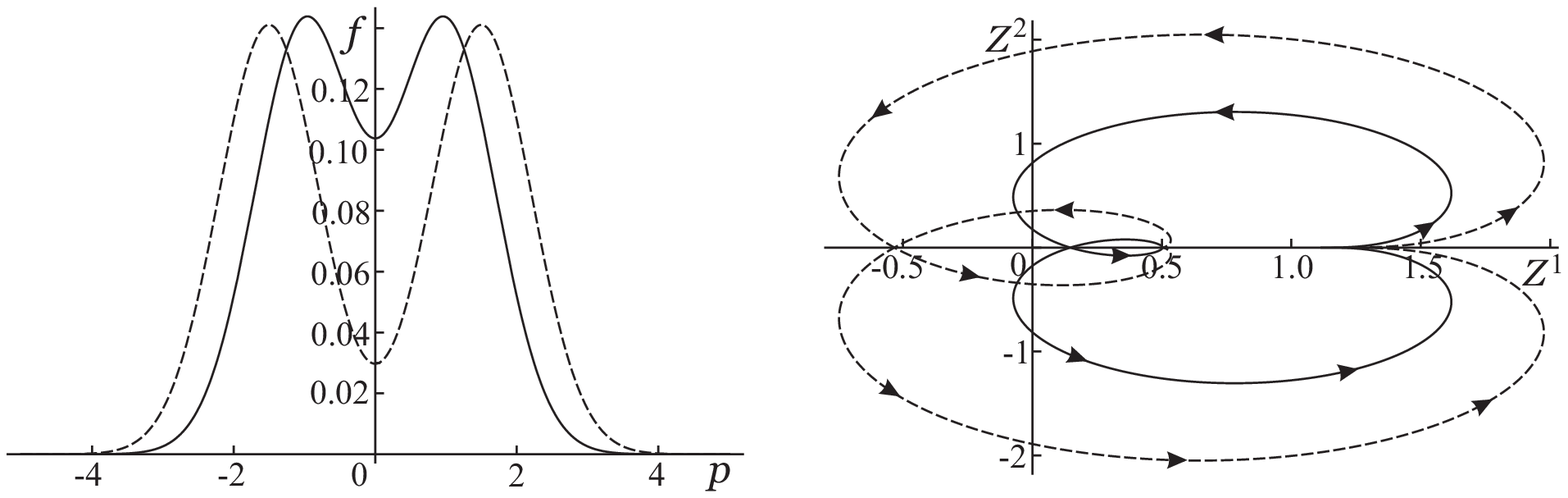}} \\[0pt]
\parbox{0.48\textwidth}{\caption{Distribution function $f(p)$ with two maxima: the solid curve corresponds to $a=1$; the dashed curve to $a=3/2$.} \label{fig:fig_1}} \hfill \parbox{0.48\textwidth}{\caption{Contours $C^+$ in the complex plane $(Z^1, Z^2)$ (notation the same as in Figure~\ref{fig:fig_1}).} \label{fig:fig_2}}
\end{center}
\end{figure}

Let us verify hyperbolicity conditions \eqref{eq:hyp-cond} for the distribution function with two maxima 
\[ f(p)=\frac{1}{4\sqrt{\pi}}  \Big(\exp\big(-(p-a)^2\big)+
   \exp\big(-(p+a)^2 \big)\Big), \quad p\in (-\infty,\infty). \]
Figure~\ref{fig:fig_1} shows a plot of this function for $a=1$ (solid curve) and $a=3/2$ (dashed). The contours $C+$ corresponding to these functions are plotted in Figure~\ref{fig:fig_2}; the direction of circulation about the contours is positive (counter-clockwise). As can be seen from the graphs, the hyperbolicity conditions \eqref{eq:hyp-cond} are fulfilled for the distribution function with closely spaced peaks and a small difference in the values of the local maxima and minima (solid curve in Figure~\ref{fig:fig_1}). Increasing the distance between the peaks of the distribution function, as well as the amplitude of local extrema, leads to an increment in the argument of the functions $\chi^\pm$ (dashed line in Figure~\ref{fig:fig_2}). This corresponds to existence of the complex characteristic roots and, consequently, the flow is unstable.

\section{Conservation form of the model}

The evolution of a smooth solution of the hyperbolic system of equations can involve a gradient catastrophe. Therefore, further description of the solutions is possible only for the class of discontinuous functions. It leads to the necessity to formulate the model in a form of conservation laws. To do this, we will present \eqref{eq:RS-Larg} in the form 
\begin{equation}  \label{eq:RS-CL}
 \begin{array}{l} \displaystyle 
   H_t+\big((p-j)H\big)_x=0, \quad p_{\lambda t}+\big((p-j)p_\lambda\big)_x=0, 
   \\[2mm] \displaystyle 
   j_t+\big(A_2-3j^2/2\big)_x=0 \quad 
   \Big(j=\int p H\,d\lambda, \quad A_2=\int p^2 H\,d\lambda, \quad H=p_\lambda f \Big).
 \end{array}
\end{equation}
The first condition in \eqref{eq:RS-CL} is the local conservation of a number of bubbles, the second condition is equivalent to the conservation of the function $f$ along the trajectory, and the last equation is the conservation law for hydrodynamic momentum $j$. Note that Eqs.\,\eqref{eq:RS-CL} are similar to conservation laws of the shallow water equations for shear flows \cite{TRC04, ChKh13}.

From conservation laws \eqref{eq:RS-CL} we obtain the Hugoniot conditions at the shock front $x=x(t)$ moving with the velocity $V=x'(t)$: 
\begin{equation}  \label{eq:shock-cond}
 \begin{array}{l} \displaystyle 
  \big[(p-j-V)H\big]=0, \quad \big[(p-j-V)p_\lambda\big]=0, \quad \big[A_2-3j^2/2-Vj\big]=0.
 \end{array}
\end{equation}
Here $[\varphi(t,x,\lambda)]=\varphi(t,x(t)+0,\lambda)-\varphi(t,x(t)-0, \lambda)$ is a jump of the function $\varphi$ at the shock front. Calculating the ratio of continuous quantities $(p-j-V)H$ and $(p-j-V)p_\lambda$, we obtain the following jump relation: $[f]=0$.

Let us show that the systems \eqref{eq:RS-Larg} and \eqref{eq:RS-CL} are equivalent for smooth solutions. Obviously, system \eqref{eq:RS-CL} is a consequence of equations \eqref{eq:RS-Larg} obtained by simple transformations. Next we show that \eqref{eq:RS-Larg} is a consequence of 
\eqref{eq:RS-CL}. From the first two equations \eqref{eq:RS-CL} we have $f_t+(p-j)f_x=0$. 
Integrating with respect to $\lambda$ the second equation in \eqref{eq:RS-CL} gives the equality 
\[ p_t+(p-j)p_x-pj_x=F(t,x), \]
where $F(t,x)$ is an arbitrary function. Let us multiply this equation by $H$ and integrate with respect to $\lambda$. Taking into account the equality 
\[ (p_t+(p-j)p_x)H=(pH)_t+((p-j)pH)_x\,, \]
which holds by virtue of \eqref{eq:RS-CL}, after integration we have 
\[ j_t+\big(A_2-3j^2/2\big)_x = n F. \]
The third equation in \eqref{eq:RS-CL} leads to the identity $F\equiv 0$, which proves equivalence of \eqref{eq:RS-Larg} and \eqref{eq:RS-CL} for smooth solutions.

\subsection{Approximate model}

To derive differential conservation laws that appro\-xi\-mate the integro-differential model \eqref{eq:RS-CL}, we divide the segment $\lambda\in [0, 1]$ into $M$ intervals $0\leq\lambda_0 <\lambda_1<...< \lambda_{M-1}<\lambda_M\leq 1$ and introduce the
variables 
\[ p_i(t,x)=p(t,x,\lambda_i) \quad h_i=p_i-p_{i-1}, \quad 
   \bar{p}_i=\frac{p_i+p_{i-1}}{2}, \quad 
   \bar{f}_i(t,x)=\frac{1}{h_i}\int\limits_{\lambda_{i-1}}^{\lambda_i} H\,d\lambda. \]
Taking into account the equality $H\,d\lambda=f\,dp$ and using a piecewise constant approxi\-ma\-tion of the distribution function 
\[ f(t,x,p)=\bar{f}_i(t,x), \quad p \in [p_{i-1},p_i] \]
we have the equalities 
\[ \int\limits_{\lambda_{i-1}}^{\lambda_i} pH\,d\lambda= 
   \bar{f}_i h_i \bar{p}_i, \quad  
   A_2= \sum\limits_{i=1}^M \int\limits_{p_{i-1}}^{p_i} p^2 f\,dp =
   \sum\limits_{i=1}^M \bigg(\frac{\bar{f}_ih_i^3}{12}+\bar{f}_i h_i \bar{p}_i^2 \bigg). \]
Next, we integrate the first two equations in \eqref{eq:RS-CL} with respect to $\lambda$ over the intervals $(\lambda_{i-1},\lambda_i)$ and use the previous formulae. As result we obtain a system of conservation laws consisting of $2M+1$ differential equations for the unknown functions $h_i(t,x)$, $\bar{f}_i(t,x)$, and $j(t,x)$: 
\begin{equation}  \label{eq:RS-CL-diff}
 \begin{array}{l} \displaystyle 
   \frac{\partial h_i}{\partial t}+ \frac{\partial}{\partial x} 
   \Big((\bar{p}_i-j)h_i\Big)=0, \quad \frac{\partial }{\partial t}
   \Big(\bar{f}_i h_i\Big)+ \frac{\partial}{\partial x}
   \Big(\big(\bar{p}_i-j\big)\bar{f}_i h_i\Big)=0, \\[3mm]\displaystyle 
   \frac{\partial j}{\partial t}+ \frac{\partial}{\partial x} \bigg(\sum\limits_{i=1}^M \Big(\frac{\bar{f}_i h_i^3}{12}+ \bar{f}_i h_i \bar{p}_i^2\Big)-\frac{3}{2}j^2\bigg)=0.
 \end{array}
\end{equation}
The quantities $p_{ci}(t,x)$ included in \eqref{eq:RS-CL-diff} are given by the formulae 
\[ \bar{p}_i=-\frac{h_i}{2}+\sum\limits_{k=1}^{i} h_k + 
   \bigg(\sum\limits_{i=1}^M  \bar{f}_i h_i \bigg)^{-1} \bigg(j+\frac{1}{2}
   \sum\limits_{i=1}^M \bar{f}_i h_i^2- \sum\limits_{i=1}^M \bar{f}_i h_i
   \sum\limits_{k=1}^i h_k\bigg). \]

To solve the differential conservation laws \eqref{eq:RS-CL-diff} numerically, one can apply Godunov type methods. In this case, due to a large number of equations in the system \eqref{eq:RS-CL-diff}, it is convenient to use central schemes \cite{NT90}, which do not require exact or approximate solution of the Riemann problem.

System \eqref{eq:RS-CL-diff} can be written in the following form 
\begin{equation}  \label{eq:Au}
  \mathbf{u}_t+A(\mathbf{u})\mathbf{u}_x=0,
\end{equation}
where $\mathbf{u}=(h_1,...,h_M,\bar{f}_1,...,\bar{f}_M,j)$ and $A(\mathbf{u})$ is a corresponding matrix. As follows from \eqref{eq:RS-CL-diff}, the functions $\bar{f}_i$ satisfy the equations $\bar{f}_{it}+(\bar{p}_i-j)\bar{f}_{ix}=0$. It means that the matrix $A(\mathbf{u})$ has the following eigenvalues: $k^*_i=\bar{p}_i-j$ ($i=1,...,M$). To find the others eigenvalues of $A$, one can use the equation 
\begin{equation}  \label{eq:chi-f-d}
  \bar{\chi}(k)=1-n+(k+j)^2\sum\limits_{i=1}^M \Big(\frac{1}{q_{i-1}-k}-\frac{1}{q_i-k}\Big)\bar{f}_i =0
\end{equation}
where $\bar{\chi}(k)$ is a discrete analogue of the characteristic function \eqref{eq:chi-f}, $q_i=p_i-j$ and the moments $n$ and $j$ are given by formulae 
\[ n=\sum\limits_{i=1}^M \bar{f}_i h_i, \quad 
   j=\sum\limits_{i=1}^M \bar{p}_i \bar{f}_i h_i\,. 
\]
In the hyperbolic case Eq.~\eqref{eq:chi-f-d} has $M+1$ real roots $k_i$ ($i=1,...,M+1$) in the interval $(p_0-j,p_M-j)$.

\begin{figure}[htb]
\begin{center}
\resizebox{0.55\textwidth}{!}{\includegraphics{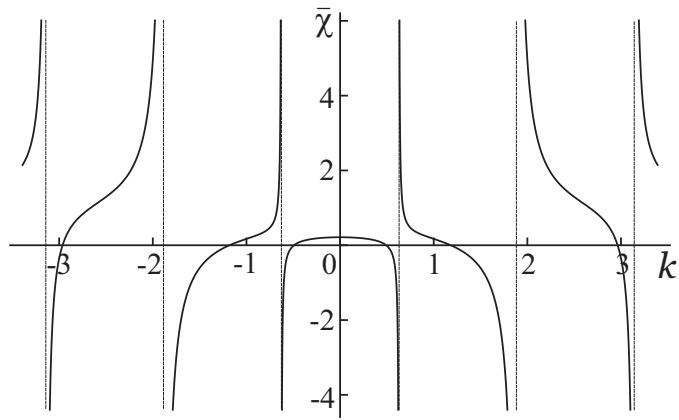}} \\[0pt]
\end{center}
\caption{Graph of the function $\bar{\protect\chi}(k)$ for $M=5$.}
\label{fig:fig_3}
\end{figure}

Let the values $\bar{f}_i$ satisfy the inequalities 
\[ \bar{f}_1<\bar{f}_2<...<\bar{f}_l, \quad \bar{f}_M<\bar{f}_{M-1}<...<\bar{f}_l, \]
that correspond to an approximation of the distribution function with a single maximum. Then the equation $\bar{\chi}(k)=0$ has a unique root $k=k_i$ on the each interval $k\in (q_{i-1},q_i)$, $i=1,...,l-1,l+1,...,M$. Indeed, function $\bar{\chi}(k)$ is monotonic and varies from $-\infty$ to $\infty$ on the intervals $(q_{i-1},q_i)$ for $i=1,...,l-1$, and from $\infty$ to $-\infty$ for $i=l+1,...,M$. Consequently, the system \eqref{eq:RS-CL-diff} is hyperbolic if the equation \eqref{eq:chi-f-d} has two roots in the interval $(q_{l-1},q_l)$. Since $\chi(-j)=1-n>0$ and $\bar{\chi}(k)$ tends to $-\infty$ as $k\to q_{l-1}$ (or $k\to q_l$), then the inequalities 
\[ p_{l-1}<0<p_l \quad (q_{l-1}<-j<q_l) \]
are the sufficient conditions for hyperbolicity of the system \eqref{eq:RS-CL-diff}.

The typical form of the function $\bar{\chi}(k)$ for $M=5$ is shown in Figure~\ref{fig:fig_3}. This graphic is obtained for the piecewise constant approximation of the distribution function with bounded support: $f(p)=(\cos p+1)/8$, $p \in (-\pi,\pi)$. Note that the integro-differential model~\eqref{eq:RS-Larg} has continuous characteristic spectrum $k=p-j$, where $p\in \mathrm{supp}\, f$.

\section{Conservation laws and reductions}

Consequence of the kinetic model \eqref{eq:RS-1D} is the following infinite chain of equations 
\begin{equation}  \label{eq:chain}
 \frac{\partial A_k}{\partial t}+ \frac{\partial A_{k+1}}{\partial x}- A_1
 \frac{\partial A_k}{\partial x}- (k+1)A_k\frac{\partial A_1}{\partial x}=0
\end{equation}
for the moments of the distribution function 
\[ A_k=\int p^k f\,dp \quad (k=0,1,2,...). \]
The chain \eqref{eq:chain} can be written as infinite series of conservation laws 
\begin{equation}  \label{eq:P-Q}
  \frac{\partial P_k}{\partial t}+\frac{\partial Q_k}{\partial x}=0,
\end{equation}
where $P_k$ and $Q_k$ are polynomials in the $A_i$ $(i=0,...,k+1)$. The first five conservation laws have the form 
\[ \begin{array}{l}\displaystyle 
    P_0=A_0, \quad Q_0=(1-A_0)A_1; \quad P_1=A_1, \quad
    Q_1=A_2-3A_2^2/2; \\[2mm]\displaystyle 
    P_2=A_2-A_1^2, \quad Q_2=A_3-3A_1A_2+2A_1^3; \\[2mm]\displaystyle
    P_3=A_3-3A_1A_2+2A_1^3, \quad Q_3=A_4-4A_1A_3-A_2^2+8A_1^2A_2-4A_1^4; 
    \\[2mm]\displaystyle 
    P_4=A_4-4A_1A_3-2A_2^2+10A_1^2A_2-5A_1^4, \\[2mm]\displaystyle 
    Q_4=A_5-5A_1A_4-4A_2A_3+14A_1^2A_3+12A_1A_2^2-30A_1^3A_2+12A_1^5.%
  \end{array} \]

According to \cite{Tesh99}, algorithm for constructing conservation laws is as follows. Function 
\[ p(t,x,\xi)=-\xi+a_1(t,x)+a_2(t,x)\xi^{-1}+a_3(t,x)\xi^{-2}+... \]
is the density of the conservation law 
\begin{equation}  \label{eq:CL-Tesh}
  p_t+\big(p^2/2-j p\big)_x=0.
\end{equation}
Coefficients $a_i$ are expressed in terms of the moments of the distribution function $A_1,...,A_i$. Let us introduce the function 
\[ \beta(\alpha)=a_1(t,x)+\alpha a_2(t,x)+\alpha^2 a_3(t,x)+... \]
Differentiation of the identity 
\[ \alpha-\frac{\alpha}{1-\alpha\beta}+ \sum\limits_{i=2}^\infty (-1)^i  
   \Big(\frac{\alpha}{1-\alpha\beta}\Big)^i A_{i-1}\equiv 0 \]
with respect to $\alpha$ at the point $\alpha=0$ allows consistently determine the coefficients $a_i$. Substituting function $p$ in the conservation law \eqref{eq:CL-Tesh} and equating to zero the expressions of the same powers $\xi$ leads to equations \eqref{eq:P-Q}. Direct calculations
show that the $k$-th conservation law \eqref{eq:P-Q} (based on previous) reduces to $k$-th chain equation \eqref{eq:chain}.

The existence of an infinite number of conservation laws is a rare property of hydro\-dynamic models. Benney equations of the long-wave theory is well-known example of the model having an infinite number of conservation laws \cite{Ben73}. Both of the models, Russo--Smereka and Benney, are generalized hyperbolic in the sense of \cite{Tesh85} and can be written in terms of the Riemann invariants, which are conserved along the characteristics.

\subsection{Finite component reductions}

Under certain assumptions about the distribution function the kinetic model \eqref{eq:RS-1D} can be reduced to a system of differential equations. The following representation of the solution 
\[ f=\sum\limits_{i=1}^M n_i(t,x)\delta(p-p_i(t,x)) \]
leads to the ``frozen'' bubbly flow 
\[ \frac{\partial n_i}{\partial t}+\frac{\partial }{\partial x}\Big((p_i-j)n_i\Big)=0, \quad   
   \frac{\partial p_i}{\partial t}+\frac{\partial }{\partial x} \Big(\frac{p_i^2}{2}-p_i j\Big)=0, \quad j=\sum\limits_{i=1}^M n_i p_i. \]
Here $\delta(p)$ is a Dirac delta function. This system of equations has imaginary characteristic roots \cite{RS96b}. 

The waterbag concept known in plasma physics \cite{David72} allows one to obtain partial solutions of the kinetic equation \eqref{eq:RS-1D} described by closed system of equations. We represent the $k$-th moment in the form~\cite{Pavlov2008} 
\[ A_k=\frac{1}{k+1}\sum\limits_{i=0}^M \varepsilon_i p_i^{k+1}, \quad
  \sum\limits_{i=0}^M \varepsilon_i=0, \]
where $\varepsilon_i$ are arbitrary constants whose sum is equal to zero. Substitution of the above introduced moments into the chain \eqref{eq:chain} yields a closed system of $M+1$ equations for the unknown functions $p_i(t,x)$: 
\begin{equation}  \label{eq:pi}
  \frac{\partial p_i}{\partial t} + (p_i-j)\frac{\partial p_i}{\partial x} 
  -p_i\frac{\partial j}{\partial x} =0, \quad j=\frac{1}{2}\sum\limits_{i=0}^M
  \varepsilon_i p_i^2\,.
\end{equation}
This system coincides with the conservation laws \eqref{eq:RS-CL-diff} only for the class of solutions $\bar{f}_i=f_i=\mathrm{const}$ (in this case \eqref{eq:RS-CL-diff} is an exact consequence of \eqref{eq:RS-1D}). Note that due to the following representation of the solution (step function): 
\[ \begin{array}{l} \displaystyle 
    f=\sum\limits_{i=1}^M f_i\big(\theta(p-p_{i-1})-\theta(p-p_i)\big)\equiv 
    \sum\limits_{i=0}^M \varepsilon_i\theta(p-p_i), \\[2mm]\displaystyle 
    \sum\limits_{i=0}^M \varepsilon_i=0, \quad f_{i+1}=\sum\limits_{k=0}^i \varepsilon_i,
   \end{array} \]
system \eqref{eq:pi} can be obtained directly from the kinetic model \eqref{eq:RS-1D}. This system of equations is hyperbolic under some conditions, which have been derived above.

The waterbag reduction \eqref{eq:pi} can be written in Riemann invariants 
\[ \begin{array}{l}\displaystyle 
    \frac{\partial r_m}{\partial t}+(k_m+j)\frac{\partial r_m}{\partial x}=0 
    \quad (m=1,...,M+1), \\[2mm]\displaystyle 
    r_m=\frac{n-1}{k_m+j}+\sum\limits_{i=1}^M f_i \ln \Big|\frac{p_i-k_m-j}{p_{i-1}-k_m-j}\Big|,
   \end{array} \]
where $k_m$ are the roots of the characteristic equation \eqref{eq:chi-f-d}. This representation  allows one to construct exact solutions in the form 
\[ r_i=r_{i0}=\mathrm{const} \quad (i\neq s), \quad r_s=r_s(k), \quad k=k_s+j. \]
These relations define functions $p_0(k),...,p_M(k)$, where $k(t,x)$ is a solution of the equation $k_t+kk_x=0$.

The existence of Riemann invariants for the system \eqref{eq:pi} follows from the fact that integro-differential equations \eqref{eq:RS-Larg} can be presented in the characteristic form \cite{Tesh99}
\[ f_t+(p-j)f_x=0, \quad R_t+(p-j)R_x=0, \]
where $f(t,x,\lambda)$ and 
\begin{equation}  \label{eq:R} 
  R(t,x,\lambda)=\frac{n(t,x)-1}{p(t,x,\lambda)}+  
  \int\frac{f(t,x,\nu)p_\nu\,d\nu}{p(t,x,\nu)-p(t,x,\lambda)} 
\end{equation}
are Riemann invariants (here function $n$ is given by formula \eqref{eq:n}).  

\subsection{Special class of solutions}

We also consider solutions of Eq.~\eqref{eq:RS-1D} in the class of functions that are piecewise continuous in the variable $p$ with a bounded support: 
\[ f=f^{1}(t,x,p)\big[\theta (p-p_{l}(t,x))-\theta (p-p_{r}(t,x))\big]. \]
Here $\theta $ is a Heaviside step function, $p_l$ and $p_r$ are the boundaries of the interval in the variable $p$, beyond which the distribution function is identically zero, and $f^{1}(t,x,p)>0$ is a non-negative function which is continuously differentiable on the set $\{(t,x,p):\ t\geq 0,\,x\in \mathbb{R},\,p\in \lbrack p_{l},p_{r}]\}$. Substitution of this ansatz  into \eqref{eq:RS-1D} yields 
\begin{equation}\label{eq:plr}
 \begin{array}{l}\displaystyle
  \frac{\partial f}{\partial t}+(p-j)\frac{\partial f}{\partial x}+
  p\frac{\partial j}{\partial x}\frac{\partial f}{\partial p}=0,\quad
  j(t,x)=\int\limits_{p_l}^{p_r}pf(t,x,p)\,dp, \\[3mm]\displaystyle
  \frac{\partial p_l}{\partial t}+(p_l-j)\frac{\partial p_l}
  {\partial x}-p_l\frac{\partial j}{\partial x}=0,\quad \frac{\partial p_r}
  {\partial t}+(p_r-j)\frac{\partial p_r}{\partial x}-p_r\frac{\partial j}{\partial x}=0,
 \end{array}
\end{equation}
where $f=f^{1}$.

The class of solutions of the kinetic model \eqref{eq:RS-1D} with a bounded support characterized by a linear relationship between the Riemann integral invariants was obtained in \cite{RTC05} on a base of the following property: if the functions $f(t,x,p)$, $p_l(t,x)$, and $p_r(t,x)$ are a solution of system \eqref{eq:RS-1D}, \eqref{eq:plr} then the quantity $R(t,x,p)$, which has been defined above by \eqref{eq:R} in semi-Lagrangian coordinates, satisfies the equation
\[ R_t+(p-j)R_x+pj_xR_p=0. \]
Linear relationship between the invariants $R$ and $f$ 
\begin{equation}  \label{eq:lin_R_f}
  R+f\pi\cot(\mu\pi)-b=0
\end{equation}
($\mu$ and $b$ are constants) leads to a special class of solutions \cite{RTC05}. In this case the distribution function $f$ has the form 
\begin{equation}  \label{eq:Russo-fsp}
  f=\frac{\sin(\mu\pi)}{\pi} \bigg(\frac{1-\mu b(p_r-p_l)}{p}+b\bigg) \bigg(\frac{p-p_l}{p_r-p}\bigg)^{\mu}
\end{equation}
and the functions $p_l$ and $p_r$ satisfy to the system \eqref{eq:plr} with the following first moment 
\begin{equation}  \label{eq:j-sp}
  j(p_l,p_r)=\mu (p_r-p_l)\Big(1+\frac{1+\mu}{2}b p_l+\frac{1-\mu}{2}b p_r\Big).
\end{equation}

Description of discontinuous solutions from this special class is given on a base of closed system of conservation laws 
\begin{equation}  \label{eq:RS-CL-sp}
  n_t+\big((1-n)j\big)_x=0, \quad j_t+\big(A_2(n,j)-3j^2/2\big)_x=0.
\end{equation}
By virtue of \eqref{eq:Russo-fsp} the moments $n$, $j$, and $A_2$ are expressed in terms of $p_l$ and $p_r$ by \eqref{eq:j-sp} and formulae 
\[ \begin{array}{l} \displaystyle 
    n=1-\bigg(\frac{p_l}{p_r}\bigg)^\mu \Big(1-\mu b(p_r-p_l)\Big), \\[3mm]\displaystyle 
    A_2=\mu (p_r-p_l)\Big(\frac{1+\mu}{2}p_r+ \frac{1-\mu}{2}p_l+
    \frac{1-\mu^2}{3}b(p_r-p_l)^2+b p_l p_r \Big).
   \end{array} \]
Below, these conservation laws will be used to perform numerical simulation of disconti\-nuous bubbly flows from special class of solutions as well as to test results obtained on the base of general model \eqref{eq:RS-CL-diff}.

{\sf Remark.} In the general case relation \eqref{eq:lin_R_f} is not fulfilled at the shock front. It means we cannot use conservative form \eqref{eq:RS-CL-sp} for the description of discontinuous flows. Only the general system of conservation laws \eqref{eq:RS-CL-diff} can be applied for this purpose. Nevertheless, modelling of discontinuous flows with small
amplitude jumps $\delta=[n]$ is possible on the base of simplified equations \eqref{eq:RS-CL-sp}. Following \cite{LT00, TRC04}, one can show that $[R]=O(\delta^2)$ and, consequently, Hugoniot conditions 
\[ \big[(p-j-V)H\big]=0, \quad \big[R+f\pi\cot(\mu\pi)-b\big]=0, \quad 
   \big[A_2-3j^2/2-Vj\big]=0, \]
which performed for a special class of solutions, coincide with relations \eqref{eq:shock-cond} if the second-order terms in $\delta$ are neglected. Notice that the previous jump conditions are consequence of the following conservative form 
\[ \begin{array}{l} \displaystyle 
    H_t+\big((p-j)H\big)_x=0, \quad j_t+(A_2-3j^2/2)_x=0, \\[2mm]\displaystyle \big((R+f\pi\cot(\mu\pi)-b)H\big)_t+\big((R+f\pi\cot(\mu\pi)-b)(p-j)H\big)_x=0
   \end{array} \]
of the system \eqref{eq:RS-Larg}.

\subsection{Fluid dynamic limit}

Let the distribution function has the form 
\[ f=\frac{n}{\sqrt{2\pi T}}\exp\Big(-\frac{(p-\bar{p})^2}{2T}\Big). \]
Calculation of moments of this distribution function  
\[ A_0=n, \quad A_1=j=n\bar{p}, \quad A_2=(\bar{p}^2+T)n, \quad 
   A_3=(\bar{p}^2+3T)n\bar{p} \]
and their substitution into the first three conservation laws \eqref{eq:P-Q} leads to the closed system of equations for the unknown functions $n(t,x)$, $\bar{p}(t,x)$ and $T(t,x)$: 
\begin{equation} \label{eq:n-p-T}
 \begin{array}{l}\displaystyle 
  n_t+(\bar{u}n)_x=0, \quad (n\bar{p})_t+(nT+n\bar{p}\bar{u}-n^2\bar{p}^2/2)_x=0, \\[2mm]\displaystyle 
  (nT+n\bar{p}\bar{u})_t+\big((3T+(1-2n)\bar{p}^2)n\bar{u}\big)_x=0,
 \end{array}
\end{equation}
where $\bar{u}=(1-n)\bar{p}$. This system of equations, obtained in \cite{RS96b}, is hyperbolic for $T>n\bar{p}^2/3$. In fact, let us rewrite \eqref{eq:n-p-T} in the form \eqref{eq:Au}, where $\mathbf{u}=(n,\bar{p},T)$. The eigenvalues of $A(\mathbf{u})$ are 
\[ k_0=(1-n)\bar{p}, \quad k_{1,2}=(1-2n)\bar{p}\pm\sqrt{(1-n)(3T-n\bar{p}^2)}. \]
Taking $T=h_1^2/12$, $\bar{p}=\bar{p}_1$, and $n=\bar{f}_1 h_1$ it is easy to see that the system \eqref{eq:n-p-T} coincides with differential approximation \eqref{eq:RS-CL-diff} for $M=1$. Note that the eigenvalues $k_{1,2}$ are the roots of the characteristic equation \eqref{eq:chi-f-d}.

\section{Modified Benney hydrodynamic chain}

For any given function $j(t,x)$ the Liouville equation 
\begin{equation} \label{eq:KHC-a}
  f_t+(p-j)f_x+pj_x f_p=0  
\end{equation}
admits a special solution which has an asymptotic expansion ($p\rightarrow \infty $) 
\begin{equation} \label{eq:KHC-e}
  f=\frac{1}{p}+\frac{B_{0}}{p^{2}}+\frac{B_{1}}{p^{3}}+\frac{B_{2}}{p^{4}}+...
\end{equation}
Of course, a solution of linear equation \eqref{eq:KHC-a} is determined up to an arbitrary function of a single variable $\Phi (f(t,x,p))$. However, Russo--Smereka kinetic model \eqref{eq:RS-1D} is a nonlinear integro-differential equation in partial derivatives, because the function $j(t,x)$ is not given, but is determined by the integral 
\begin{equation} \label{eq:KHC-jey}
  j=\int p\Phi (f(t,x,p))\,dp,  
\end{equation}
where the function $\Phi (f(t,x,p))$ is an appropriate rapidly decreasing function at infinities such that above integral makes sense (here we omit investigation of the question: how to determine this function $\Phi(f(t,x,p))$. This separate question was studied in \cite{Tesh99, LT00,
Chesn00}).

In comparison with previous Section, we renumerate moments $A_{k}\equiv B_{k-1}$ \eqref{eq:chain} according to Kupershmidt's notation (see \cite{Kuper}), i.\,e. (here we replace a general solution $f(t,x,p)$ of the Russon--Smereka kinetic model by rapidly decreasing function $\Phi (f(t,x,p))
$ at infinities that all integrals for moments make sense, see for instance  \cite{David72}, where $f(t,x,p)$ is a special solution, which has asymptotic expansion \eqref{eq:KHC-e})
\begin{equation} \label{eq:KHC-mom}
  A_{k+1}\equiv B_{k}(t,x)=\int p^{k+1}\Phi (f(t,x,p))dp,\quad 
  k=0,\pm 1,\pm 2,...  
\end{equation}
Then the Russo--Smereka kinetic model \eqref{eq:RS-1D} implies the well-known modified Benney hydrodynamic chain (here we took into account that $j\equiv B_{0}$ which follows from comparison of \eqref{eq:KHC-jey} with \eqref{eq:KHC-mom}) for $k=0$) 
\begin{equation} \label{eq:KHC-mod}
  \frac{\partial B_{k}}{\partial t}+\frac{\partial B_{k+1}}{\partial x}-B_{0}
  \frac{\partial B_{k}}{\partial x}-(k+2)B_{k}\frac{\partial B_{0}}{\partial x}=0,
  \quad k=0,\pm 1,\pm 2,...  
\end{equation}
which was derived by Kupershmidt in more wide but in a pure mathematical context (see detail in \cite{Kuper} and also in \cite{MaksKuper}). This chain is connected with the Benney hydrodynamic chain (see \cite{Ben73} and also, for instance, \cite{Gibbons, KM, Zakh}) 
\begin{equation} \label{eq:KHC-benney}
 \frac{\partial C_k}{\partial t}+\frac{\partial C_{k+1}}{\partial x}
 +kC_{k-1}\frac{\partial C_0}{\partial x}=0,\quad k=0,1,...,
\end{equation}
because all moments $C_k$ can be expressed via polynomials with respect to moments $B_0,B_1,...,B_{k-1}$. So, this is not an invertible point transformation, this is an infinite set of Miura type transformations $C_k(B_0,...,B_{k-1})$, which can be computed comparing the asymptotic
expansions ($q\rightarrow \infty $, cf. \eqref{eq:KHC-e}) 
\begin{equation}\label{eq:KHC-c}
  f=q+\frac{C_0}{q}+\frac{C_1}{q^2}+\frac{C_2}{q^3}+...=
  p\left(1+\frac{B_0}{p}+\frac{B_1}{p^2}+\frac{B_2}{p^3}+...\right)^{-1},
\end{equation}
where $q=p-B_{0}$. Indeed, the kinetic equation \eqref{eq:KHC-a} transforms into the Vlasov equation (here $\tilde{\Phi}(f(t,x,q))$ is another but also rapidly decreasing solution at infinity) 
\begin{equation} \label{eq:KHC-b}
  f_t+qf_x-f_q C_{0x}=0,\quad C_0=\int\tilde{\Phi}(f(t,x,q))\,dq
\end{equation}
under the substitution $q=p-j$, where $C_{0x}=j_t-jj_x$. Moreover, substitution 
\[ C_k=\int q^k\tilde{\Phi}(f(t,x,q))\,dq, \quad B_k=\int p^{k+1}\Phi(f(t,x,p))\,dp \]
into \eqref{eq:KHC-a}, \eqref{eq:KHC-b}, respectively, leads to the following integro-differential equations: 
\[ f=q+\int \frac{\tilde{\Phi}(f(t,x,q'))\,dq'}{q-q'}, \quad 
   \frac{1}{f}=\frac{1}{p}+\frac{1}{p}\int \frac{p'\Phi(f(t,x,p'))dp'}{p-p'}, \]
which below we shall utilize for construction of particular solutions. 

{\sf Remark}: A comparison of the right equation in previous formula and \eqref{eq:R} shows that $R=-1/f$, where we took into account \eqref{eq:n}.

In comparison with Benney hydrodynamic chain \eqref{eq:KHC-benney}, modified Benney chain \eqref{eq:KHC-mod} has infinite series of local conservation laws for both positive and negative values of discrete variable $k$ (see \cite{MaksKuper}). For instance, 
\[ \begin{array}{l} \displaystyle 
     \frac{\partial B_{-1}}{\partial t}+ \frac{\partial }{\partial x } 
     \Big(B_0(1-B_{-1})\Big)=0, \\[2mm]\displaystyle 
     \frac{\partial }{\partial t} \Big(B_{-2}(B_{-1}-1)\Big)= 
     \frac{\partial }{\partial x} \Big(B_0(B_{-1}-1)B_{-2}-
     \frac{1}{2}(B_{-1})^{2}+B_{-1}\Big),...
\end{array} \]
The computation of positive local conservation law densities can be found from an inverse asymptotic expansion to \eqref{eq:KHC-c}: 
\[ p=f-H_0-\frac{H_1}{f}-\frac{H_2}{f^2}-\frac{H_3}{f^3}-..., \]
while all negative conservation law densities can be found (see \cite{MaksKuper}) inverting another asymptotic expansion ($p\rightarrow 0$) 
\[ \frac{1}{f}=\frac{B_{-1}-1}{p}+B_{-2}+pB_{-3}+p^{2}B_{-4}+... \]
This means that negative conservation law densities $H_{-k}$ can be found iteratively by substitution ($f\rightarrow 0$) 
\[ p=(B_{-1}-1)f+H_{-2}f^{2}+H_{-3}f^{3}+... \]

\subsection{Method of Hydrodynamic Reductions}

The method of hydrodynamic reductions was established in \cite{GT} for Vlasov (collisionless Boltzmann) equation \eqref{eq:KHC-b}. In the case of the Russo--Smereka kinetic model \eqref{eq:RS-1D} the same approach is applicable. Indeed, let us consider a family of $N$ component hydrodynamic systems written via Riemann invariants (i.\,e. in a diagonal form) 
\begin{equation} \label{eq:KHC-diag}
  r_t^i=\mu^i(\mathbf{r})r_x^i,
\end{equation}
such that all moments $B_k(t,x)$ depend on these $N$ Riemann invariants $r^i(t,x)$ and the distribution function $f(t,x,p)$ depends now on Riemann invariants $r^{k}$ and the momentum $p$. In such a case, equation %
\eqref{eq:KHC-a} reduces to the remarkable L\"{o}wner equations (see \cite{GT}) 
\begin{equation}  \label{eq:KHC-rim}
  \partial_i\lambda =-\frac{p\partial_i j}{\mu^i+p-j}\lambda_p,
\end{equation}
where $\partial_i=\partial/\partial r^i$, and $\lambda (\mathbf{r},p)$ is a distribution function $f(t,x,p)$ restricted on the above family of $N$ component hydrodynamic reductions, i.\,e. we consider the special family of solutions $f(t,x,p)=\lambda (\mathbf{r}(t,x),p)$.

One can check a consistency of L\"{o}wner equations \eqref{eq:KHC-rim} for each pair of distinct indices $i$ and $k$, which imply the so called Gibbons--Tsarev system describing simultaneously all admissible functions $\mu^i(\mathbf{r})$ and the zeroth moment $B_0(\mathbf{r})$. Nevertheless, the Riemann invariants are not so appropriate coordinates for straightforward computation of these functions $\mu^i(\mathbf{r})$ and $B_0(\mathbf{r})$. By this reason, we utilize an alternative approach (see detail in \cite{algebra} and in \cite{Pavlov2008}), which is based on a
special choice of coordinates $a^k(\mathbf{r})$ such that hydrodynamic type systems \eqref{eq:KHC-diag} are written in the so called symmetric form 
\begin{equation} \label{eq:KHC-hydro}
  a_t^k+\left(\frac{(a^k)^2}{2}-a^k B_0(\mathbf{a})\right)_x=0,
\end{equation}
where the function $B_0(\mathbf{a})$ satisfies the Gibbons--Tsarev system written in the form 
\begin{equation} \label{eq:KHC-gt}
  (a^i-a^k)\partial_{ik}^2 B_0+\partial_k B_0\hat{R} 
  \partial_i B_0-\partial_i B_0\hat{R} \partial_k B_0=0, \quad i\neq k.
\end{equation}
Here $\hat{R}=\Sigma a^m\partial /\partial a^m$ is a scaling operator and nonlinear system \eqref{eq:KHC-gt} can be derived, for instance, from the zero-th conservation law 
\[ (B_0(\mathbf{a}))_t+\left(B_1(\mathbf{a})-\frac{3}{2}(B_0(\mathbf{a}))^2\right)_x=0, \]
i.\,e. functions $B_0(\mathbf{a})$ and $B_1(\mathbf{a})$ satisfy the above conservation law for $N$ component hydrodynamic type system if $B_0(\mathbf{a})$ is a solution of \eqref{eq:KHC-gt}.
Alternatively, Gibbons--Tsarev system \eqref{eq:KHC-gt} can be derived from consistency of hydrodynamic reductions \eqref{eq:KHC-hydro} and their generating function of conservation laws \eqref{eq:CL-Tesh}, which follows from \eqref{eq:KHC-a} under the semi-Lagrange transformation $f(t,x,p)\rightarrow p(t,x,f)$. Indeed, taking into account that in such a case $j=B_{0}(\mathbf{a})$ and $p(\mathbf{r},f)$, one can obtain the L\"{o}wner equations written in the symmetric form 
\begin{equation} \label{eq:KHC-lowner}
  \partial_i p=p\frac{\partial_i B_0(\mathbf{a})}{a^i-p}
  \left(\sum\limits_{m=1}^N\frac{\partial_m B_0(\mathbf{a})}{a^m-p}-1\right)^{-1},  
\end{equation}
where $\partial_k\equiv \partial /\partial a^k$. Then the compatibility conditions $\partial_k(\partial_i p)=\partial_i(\partial_k p)$ yield again Gibbons--Tsarev system \eqref{eq:KHC-gt}.

The Gibbons--Tsarev system possesses infinitely many solutions parametrized by $N$ arbitrary functions of a single variable for any natural number $N$. At this moment we can construct just multi-parametric families of solutions (see detail in \cite{MaksKuper}). Here we show simplest reductions can be found.

Suppose that $B_0(\mathbf{a})=\Sigma b_m(a^m)$, where $b_k(z)$ are unknown functions. Substitution this ansatz into \eqref{eq:KHC-gt} implies 
\begin{equation}  \label{eq:KHC-bi}
  b'_k=\epsilon_k (a^k)^{\epsilon -1},
\end{equation}
where $\epsilon$ and $\epsilon_k$ are arbitrary parameters. If $\epsilon\neq 0$, then hydrodynamic reductions \eqref{eq:KHC-hydro} take the form 
\begin{equation}  \label{eq:KHC-puiseux}
  a_t^k+\left(\frac{(a^k)^2}{2}-\frac{a^k}{\epsilon }\sum\limits_{m=1}^N
  \epsilon_k(a^k)^{\epsilon }\right)_x=0.
\end{equation}
If $\epsilon =0$, then 
\begin{equation}  \label{eq:KHC-log}
 a_t^k+\left(\frac{(a^k)^2}{2}-a^k\sum\limits_{m=1}^N \epsilon_k \ln a^k\right)_x=0.
\end{equation}

Hydrodynamic type systems \eqref{eq:KHC-puiseux}, \eqref{eq:KHC-log} can be integrated by the generalized hodograph method (see \cite{Tsar}). Thus, one can construct a general solution $a^{k}(t,x)$ for each of these systems in implicit form. This means that simultaneously the function $B_0(\mathbf{a}(t,x))$ can be found and corresponding solution of L\"{o}wner equations %
\eqref{eq:KHC-lowner} too. For instance, substitution \eqref{eq:KHC-bi} into \eqref{eq:KHC-lowner} leads to 
\begin{equation} \label{eq:KHC-pi}
  \partial _{k}p=p\frac{\epsilon_k(a^k)^{\epsilon -1}}{a^k-p}
  \left(\sum\limits_{m=1}^N\frac{\epsilon_m(a^m)^{\epsilon}}{a^m-p}-1\right)^{-1},  
\end{equation}
which can be integrated using hypergeometric functions. If $\epsilon $ are integers, then $f(\mathbf{a},p)$ can be expressed via elementary functions. For example, if $\epsilon =1$, integration of \eqref{eq:KHC-pi} implies 
\begin{equation} \label{eq:KHC-int}
  f(\mathbf{a},p)=p^{\Sigma \epsilon_m-1}\prod\limits_{m=1}^N(p-a^m)^{-\epsilon_m},  
\end{equation}
which has precisely the asymptotic expansion \eqref{eq:KHC-e}. Thus, corresponding particular solution of Russo--Smereka kinetic equation \eqref{eq:RS-1D} depends on $N$ arbitrary functions of a single variable, which contain in a general solution $a^{k}(t,x)$ of hydrodynamic reduction 
\eqref{eq:KHC-puiseux} for $\epsilon =1$. This means that in such a case \eqref{eq:KHC-int} contains this functional freedom changing independent field variables $a^{k}$ to functions $a^{k}(t,x)$ according to the generalized hodograph method allowing to solve \eqref{eq:KHC-puiseux},
i.\,e. we finally obtain a particular solution for the distribution function $f(t,x,p)$ given in implicit form
\[ f(\mathbf{a}(t,x),p)=p^{\Sigma \epsilon_m-1}\prod\limits_{m=1}^N(p-a^m(t,x))^{-\epsilon_m}. \]

Choosing another values of the parameter $\epsilon$ or another solutions of Gibbons--Tsarev system \eqref{eq:KHC-gt}, one can find much more complicated solutions $f(\mathbf{a},p)$ of L\"{o}wner equations \eqref{eq:KHC-lowner}. Solving corresponding hydrodynamic type systems \eqref{eq:KHC-hydro}, one can find new solutions $f(\mathbf{a}(t,x),p)$ of the Russo--Smereka kinetic model.

\section{Numerical results} 

In this section we present some results related to the numerical modelling of bubbly flows. We implement here the Nessyahu--Tadmor second-order central scheme \cite{NT90}
\begin{equation}\label{eq:predictor}
\begin{array}{l}\displaystyle
  \bu_{j}^{n+1/2}=\bu_{j}^{n}-\Lambda \bbf'_j/2, \quad\quad 
  (\Lambda=\Delta t/\Delta x)  \\[2mm]\displaystyle
  \bu_{j+1/2}^{n+1}=(\bu_j^n+\bu_{j+1}^n)/2+ 
  (\bu'_j-\bu'_{j+1})/8-\Lambda\big(\bbf(\bu_{j+1}^{n+1/2})-\bbf(\bu_{j}^{n+1/2})\big).
 \end{array}
\end{equation}
This scheme approximate the system of conservation laws of the form
\[ \bu_t+(\bbf(\bu))_x=0 \]
including models \eqref{eq:RS-CL-diff} and \eqref{eq:RS-CL-sp}. Here $\Delta x$ is the spatial grid spacing, while $\Delta t$ is the time-step satisfying the Courant condition, and $\bu_j^n=\bu(t^n,x_j)$. The calculation domain on the $x$ axis is divided into $N$ cells, the cell centres are denoted by $x_j$. Values $\bu'_j/\Delta x$ and $\bbf'_j/\Delta x$ are approximations of the first-order derivatives with respect to $x$, calculated according to the ``ENO limiter'' procedure \cite{Harten87}. At $t=0$ the initial data $\bu_j^0$ are specified. The boundary conditions $u(t^n,x_{1-j}) = u(t^n,x_1)$ and $u(t^n,x_{N+j}) = u(t^n,x_N)$ are used, which allow calculations to be performed until the initial perturbations reach the boundaries of the computation domain. Scheme \eqref{eq:predictor} does not require exact or approximate solution of the Riemann problem that is very convenient in our case, due to a large number of equations in the system \eqref{eq:RS-CL-diff}.

\begin{figure}[htb] 
\begin{center}
\resizebox{1\textwidth}{!}{\includegraphics{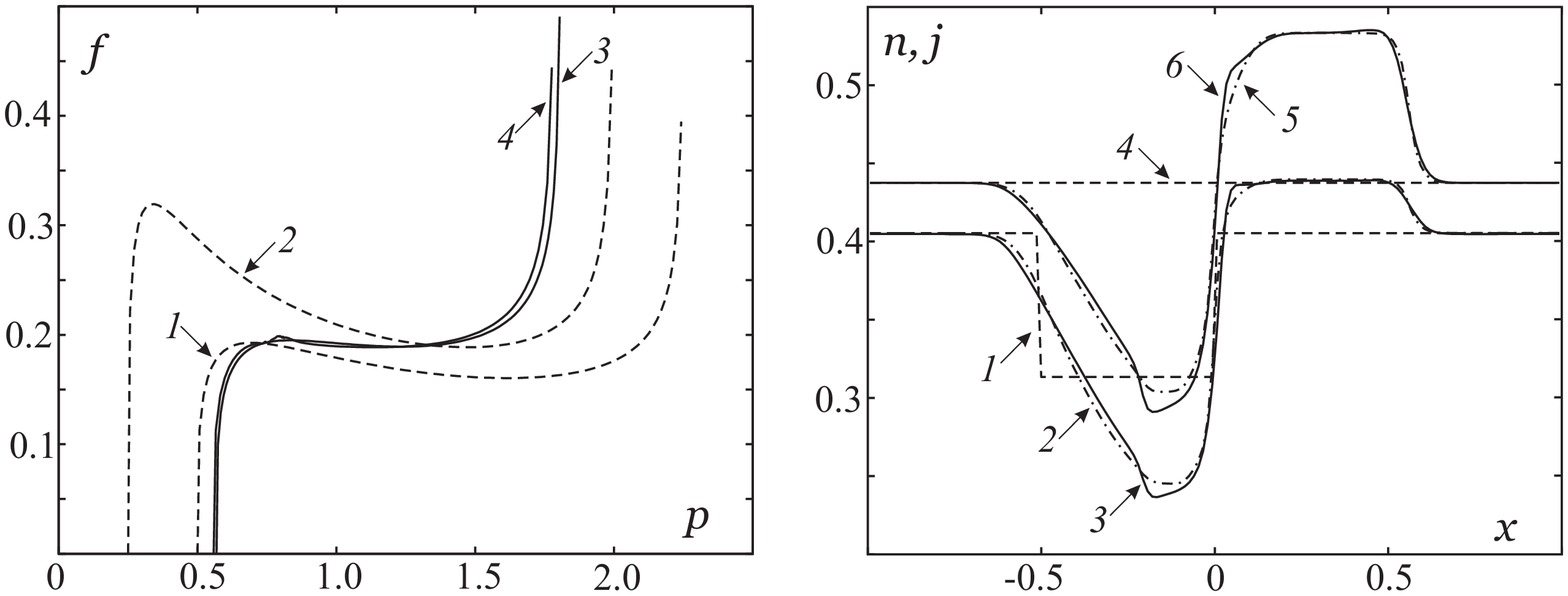}} \\[0pt]
\parbox{0.48\textwidth}{\caption{Distribution function $f$ at $t=0$ (curves {\it 1} and {\it 2}) and $t=1$ (curves {\it 3} and {\it 4}): {\it 1} --- $x\in (-0{.}5,0)$; {\it 2} --- $x\notin (-0{.}5,0)$; curves {\it 3} and {\it 4} refer to calculation using models \eqref{eq:RS-CL-sp} and \eqref{eq:RS-CL-diff} at $x=-0{.}2$.} \label{fig:fig_4}} \hfill
\parbox{0.48\textwidth}{\caption{Density $n$ (curves {\it 1 -- 3}) and moment $j$ (curves {\it 4 -- 6}): {\it 1} and {\it 4} --- $t=0$; {\it 2} and {\it 5} --- $t=1$ calculation using model \eqref{eq:RS-CL-sp}; {\it 3} and {\it 6} --- $t=1$ calculation using model \eqref{eq:RS-CL-diff}.} \label{fig:fig_5}}
\end{center} 
\end{figure}

Let us perform a comparison between the numerical solutions obtained for differential approximation \eqref{eq:RS-CL-diff} and for system \eqref{eq:RS-CL-sp} defining solutions from this special class. 
At $t=0$ we define the function $f$ by the formula \eqref{eq:Russo-fsp}, where $p_l(x)=0{.}5$ and $p_r(x)=2{.}25$ if $x\in (-0{.}5,0)$, otherwise $p_l=0{.}25$ and $p_r=2$. Initial distribution function $f(x,p)$ and moments $n(x)$ and $j(x)$ are shown in Figures \ref{fig:fig_4} and \ref{fig:fig_5} by dashed lines. 

We calculate the solution in the domain $x\in [-1,1]$ for $N=150$ and $M=120$. To start calculations on the base of the model \eqref{eq:RS-CL-diff} we specify the unknown quantities 
\[ (h_1,...,h_M, \bar{f}_1 h_1,...,\bar{f}_M h_M, j) \] 
at $t=0$. Here $h_i=(p_r-p_l)/M$; $\bar{f}_i$ is the mean value of the function $f(0,x,p)$ in the interval $(p_{i-1},p_i)$, where $p_i=p_l+ih_i$. The following parameters for the special class of solutions are chosen $\mu=0{.6}$, $b=0$. A sufficiently detailed resolution in the variable $p$ is needed due to the rapid change of the distribution function in a special class of solutions in the neighbourhood of the point $p_l$ and $p_r$. Figures \ref{fig:fig_4} and \ref{fig:fig_5} (solid lines) show the results of computation at $t=1$. This test confirms that the calculations of discontinuous solutions obtained by the ``multilayer'' approximation \eqref{eq:RS-CL-diff} and by the system for special solutions \eqref{eq:RS-CL-sp} give close results, at least for small amplitude jumps.

\subsection{Kinetic roll-over} 

In the theory of quasineutral collisionless plasma flows the following result is known \cite{Gur-Pit, KhCh11}: during evolution the kinetic roll-over of the distribution function is possible (the formation of two peaks of the distribution function which original\-ly had a single peak). Let us establish a similar property for the considered kinetic model for a bubbly flow.

\begin{figure}[htb] 
\begin{center}
\resizebox{1\textwidth}{!}{\includegraphics{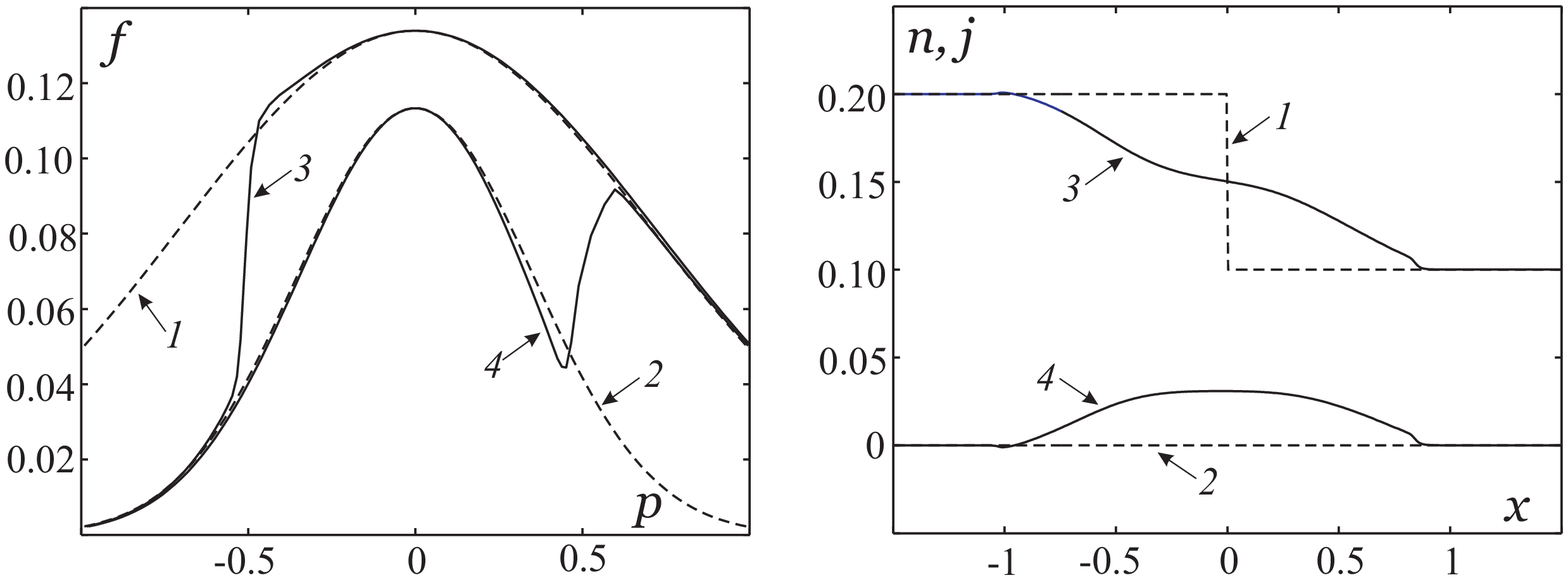}} \\[0pt]
\parbox{0.48\textwidth}{\caption{Distribution function $f$ at $t=0$ (curve {\it 1} --- $x<0$, curve {\it 2} --- $x>0$) and at $t=2$ (curve {\it 3} ---  $x=-0.5$, curve {\it 4} ---  $x=0.5$).} 
\label{fig:fig_6}} \hfill
\parbox{0.48\textwidth}{\caption{Moments $n$ and $j$ at $t=0$ (curves {\it 1, 2}) and at $t=2$ (curves {\it 3} and {\it 4}).} \label{fig:fig_7}}
\end{center} 
\end{figure}

Suppose that at $t=0$, the bubbly flow in the half-space $x<0$ is defined by a distribution function $f=f_l(p)$, and in the half-space $x>0$, by a distribution function $f=f_r(p)$. In the interval  $p\in (p_l,p_r)$, we choose the functions $f_l$ and $f_r$ as follows:
\[ f_l(p)=\frac{\exp(-p^2)}{5\sqrt{\pi} {\rm erf}(1)}, \quad
   f_r(p)=\frac{\exp(-4p^2)}{5\sqrt{\pi} {\rm erf}(2)} \quad\quad
   (p_r=-p_l=1) \]
(curves {\it 1} and {\it 2} in Figure~\ref{fig:fig_6}); outside this interval $f_l=f_r=0$. 

Distribution function at $t=2$ for $x=-0.5$ and $x=0.5$ is shown in Figure~\ref{fig:fig_6} (curves {\it 3} and {\it 4}, correspondingly). It can be seen that the distribution function for fluid more saturated by bubbles (curves 1 and 3) has changed in the minimum of $p$. At the same time, the distribution function of the less saturated fluid (curves {\it 2} and {\it 4}) has qualitative changes due to kinetic roll-over. In the vicinity of $x=0$ (point of discontinuity at $t=0$) the distribution function $f$ for $t>0$  has the following form: $f\to f_l(p)$ for $p<0$ and $f\to f_r(p)$ for $p>0$. Note that the formation of two local peaks of the distribution function usually leads to loss of the hyperbolicity of the kinetic model. Figure~\ref{fig:fig_7} shows plots of the hydrodynamic moments at $t=2$ (solid curves) and $t=0$ (dotted). In the calculations we used the following resolution $N=400$ and $M=100$. Increase or decrease the number of nodes has no significant effect on the numerical results. Solution of the Riemann problem with other initial data (functions $f_l(p)$ and $f_r(p)$ have one maximum and $\int pf_l\,dp>\int pf_r\,dp$) has a similar form.

\section{Conclusion}

Conservation form of the Russo--Smereka kinetic equation \eqref{eq:RS-1D} is proposed and differential conservation laws \eqref{eq:RS-CL-diff} approximating the model are derived. Some known fluid dynamic limits (equations \eqref{eq:n-p-T}, obtained under the assumption of local thermodynamic equilibrium, and ``waterbag'' reduction \eqref{eq:pi}) are special cases of this system of equations. An example of verification of the hyperbolicity conditions of the kinetic model is given. It is established that the distribution function with two peaks (Figures~\ref{fig:fig_1} and \ref{fig:fig_2}) leads to instability for a bubbly flow. Conservation laws \eqref{eq:RS-CL-diff} and \eqref{eq:RS-CL-sp} are used to perform numerical calculations of wave propagation in a bubbly flow initiated by discontinuous Cauchy data. It is shown the correspondence of the numerical results in a framework of the proposed approximation and in the special class of solutions characterized by a linear relationship between the Riemann integral invariants (Figures~\ref{fig:fig_4} and \ref{fig:fig_5}). The effect of the kinetic roll-over of the distribution function is demonstrated (Figure~\ref{fig:fig_6}).

\section*{Acknowledgements}

Authors thank S.\,L. Gavrilyuk for his stimulating and clarifying discussions.

AAC's work was supported by the Russian Foundation for Basic Research (grant No. 13-01-00249) and Integration Project of SB RAS No.\,30. MVP's work was partially supported by the RF Government grant 11.G34.31.0005 and by the grant of RAS ``Fundamental Problems of Nonlinear Dynamics''.

\renewcommand\baselinestretch{1}

\end{document}